\documentclass[]{spie}  %>>> use for US letter paper
\usepackage{amsmath,amsfonts,amssymb}
\usepackage{graphicx}
\usepackage[colorlinks=true, allcolors=blue]{hyperref}
\usepackage{siunitx}
\usepackage{comment}

\title{Digital frequency multiplexing with sub-Kelvin SQUIDs}

\author[a]{Amy E. Lowitz}
\author[a,b]{Amy N. Bender}
\author[c]{Matthew A. Dobbs}
\author[c]{Adam J. Gilbert}
\affil[a]{Kavli Institute for Cosmological Physics, University of Chicago, 5640 S. Ellis Avenue, Chicago, IL, USA}
\affil[b]{Argonne National Laboratory, High-Energy Physics Division, 9700 S. Cass Ave., Argonne, IL, USA}
\affil[c]{Dept. of Physics, McGill Univ., 3600 Rue University, Montreal, QC, Canada}

\authorinfo{Further author information: (Send correspondence to A.L.)\\
A.L.: E-mail: lowitz@uchicago.edu}

\begin{document} 
\maketitle

\begin{abstract}
Digital frequency multiplexing (dfMux) is a readout architecture for transition edge sensor-based detector arrays and is used on telescopes including SPT-3G, POLARBEAR-2, and LiteBIRD. Here, we present recent progress and plans for development of a sub-Kelvin SQUID architecture for digital frequency multiplexed bolometers. This scheme moves the SQUID from the \SI{4}{K} stage to the \SI{250}{mK} stage, adjacent to the bolometers. Operating the SQUID on the detector stage may offer lower noise and greater scalability. Electrical performance will be improved as a result of decreased wiring length and reduced parasitics, allowing for higher multiplexing factors and lower bolometer R$_{\textrm{normal}}$. These performance improvements will enable ultra-large focal planes for future instruments such as CMB-S4.
\end{abstract}


\section{INTRODUCTION}
\label{sec:intro}  % \label{} allows reference to this section

Digital frequency multiplexing (dfMux) is a method of reading out many transition edge sensor (TES) bolometers with a single pair of wires, which reduces wiring complexity and thermal load. Multiplexing factors of up to 68x have been demonstrated on-sky with this method\cite{anderson}.  Higher multiplexing factors are in development; SRON has demonstrated frequency division multiplexing with a 176x multiplexing factor in the laboratory\cite{SRON}, and an upgrade of the dfMux architecture to 128x multiplexing is in active development.  

One of the key challenges facing the current dfMux architecture is parasitic impedance, especially from the wiring between the inductive-capacitive (LC) filters on the \SI{250}{mK} detector stage and the SQUID (superconducting quantum interference device) amplifier on the \SI{4}{K} stage.  Parasitic inductance in this portion of the system is the dominant source of cross-talk in the cold readout (see section \ref{sec:motivation}).  Reducing parasitics in this part of the cold readout will reduce cross-talk and increase scalability.  Reduced parasitics will have the additional benefit of enabling operation of bolometers with lower normal state resistance which enables bolometer operation with lower bias voltage and lower noise.  

This work achieves significant reduction in the stray inductance between the LC filters and the SQUID by moving the SQUID to the \SI{250}{mK} stage, adjacent to the LC filter chip.  With this modified dfMux architecture, we demonstrate operation of three types of SQUID arrays at \SI{250}{mK} (section \ref{sec:SQUID}), operation of bolometers with a \SI{250}{mK} SQUID array (section \ref{sec:bolo}), and substantial reduction of parasitic inductance in the cold readout system (section \ref{sec:capTerm}).  Additionally, plans to address an in-band resonance are discussed (section \ref{sec:resTerm}).

\section{DIGITAL FREQUENCY MULTIPLEXING}

\begin{figure} [ht]
		\begin{center}
			\begin{tabular}{c} %% tabular useful for creating an array of images 
				\includegraphics[width=\textwidth]{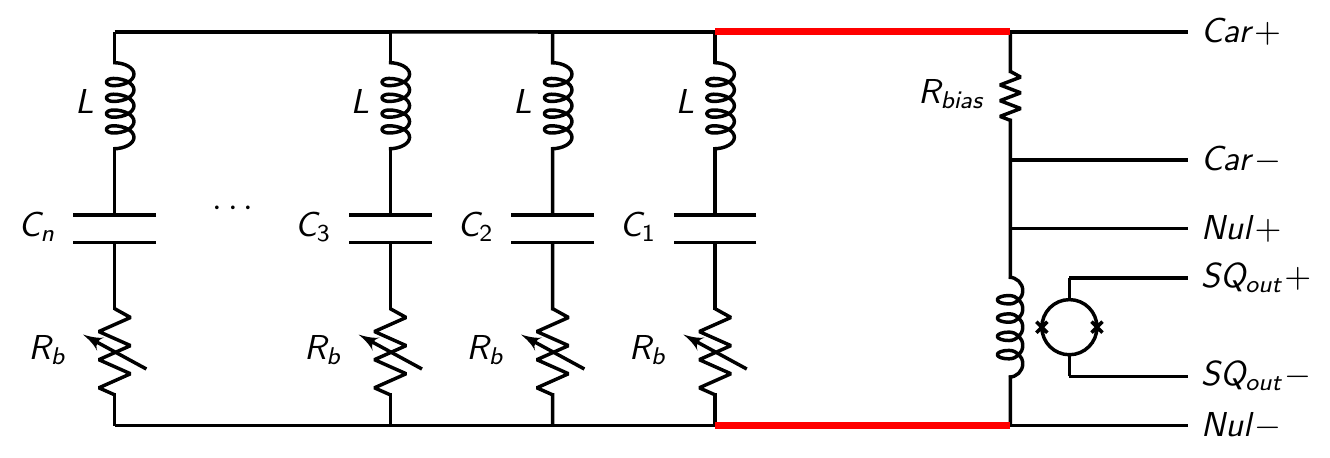}
			\end{tabular}
		\end{center}
		\caption[example] 
		{ \label{fig:dfMux} 
			(Color online) Schematic of the cold portion of the dfMux readout system.  Red indicates wiring between the LC chip and SQUID, as referenced in section \ref{sec:motivation}.  L and C$_n$ refer to the planar superconducting inductors and capacitors that make up the LC filters on the LC chip.  R$_b$ refers to the variable TES bolometer resistance.
		}
		\end{figure}

		Digital frequency multiplexing is currently used by the South Pole Telescope third-generation camera (SPT-3G)\cite{anderson}, POLARBEAR-2 (PB2)\cite{darcy}, and others\cite{ebex,litebird}. In dfMux, an RLC resonator is constructed by connecting the bolometer (R) in series with an inductor-capacitor (LC) resonator or filter.  These LC resonators consist of a lumped, coiled, superconducting inductor in series with a lumped, interdigitated, superconducting capacitor\cite{rotermund,hattori}.  An array of these resonators (68 for SPT-3G\cite{bender} and 40 for PB2\cite{darcy}) are lithographed as a single layer on a silicon chip (referred to hereafter as an ``LC chip") which is mounted on a custom PCB (referred to hereafter as an ``LC board").  Each capacitor is slightly different, so that each RLC resonator has a different center frequency.  This assigns each bolometer a distinct readout frequency.  Many RLC resonators (68, in the case of SPT-3G) are connected in parallel and voltage-biased with AC tones at each RLC frequency, summed together to produce a single combined `carrier comb' which is passed to the bolometers and LC filters in parallel with a small bias resistor. The combined signal from the bolometers is amplified with a DC SQUID array. Baseband feedback is used to null the signal at the SQUID input to improve dynamic range and allow for higher multiplexing factors\cite{deHaan}.  This ``nuller" tone is the science signal.  Figure \ref{fig:dfMux} shows a schematic of the cold portion of the readout circuit.  For a complete description of the dfMux architecture, see Bender et. al. (2014)\cite{bender}.

\subsection{Motivation}
\label{sec:motivation}
		
		The standard dfMux architecture has three primary sources of cross-talk.  First, mutual inductance between physically adjacent LC resonators on the LC chip is a source of cross-talk.  Changes in the resistance of the $i$th bolometer modulate the voltage across the $i$th inductor.  Supposing the $j$th inductor is physically near the $i$th inductor, this will induce a small voltage modulation in the $j$th inductor, proportional to the mutual inductance between L$_i$ and L$_j$, resulting in cross-talk between the $i$th and $j$th readout channels.  Second, resonance overlap between frequency-adjacent resonators is a source of cross-talk.  At any finite resonator frequency spacing, there will be some overlap between nearby resonators.  One result of this overlap is that amplitude modulation of the $i$th resonator (due to sky signal modulation of the bolometer resistance, for instance) results in a small amount of amplitude and phase modulation in the ($i\pm n$)th carrier frequency.  Finally, impedance of the wiring between the LC chip and the SQUID is the third source of cross-talk.  In the ideal case, wiring between the LC chip and the SQUID (marked in red in Figure \ref{fig:dfMux}) would have no impedance.  However in a real system wiring always has some non-zero impedance.  Impedance in these wires creates a voltage divider such that as the resistance of a bolometer is modulated, the ratio of the voltage divider is likewise modulated.  As a result, resistance modulation in one bolometer produces a modulation in the bias voltage supplied to all other bolometers, proportional to the wiring impedance.  The wiring impedance voltage divider effect is the dominant source of cross-talk in the standard dfMux architecture.  Addressing this source of cross-talk is one of the key goals of this work.  For a more detailed discussion of all three cross-talk mechanisms, see Dobbs et. al. (2012)\cite{dobbs}.
		
		Reducing the wiring impedance between the LC chip and the SQUID offers additional benefits.  The constraint on frequency spacing of the carrier comb is set by the requirements on cross-talk.  By significantly reducing the voltage divider source of cross-talk, more resonators could be read out in the same bandwidth without exceeding the tolerance for cross-talk in the dfMux system as a whole.  Denser frequency packing is essential to improving scalability of dfMux and is a key step on the path towards 128x dfMux multiplexing, and beyond.  Finally, decreased parasitics will allow for operation of bolometers with lower normal resistance (R$_{\textrm{normal}}$).  Lower R$_{\textrm{normal}}$ bolometers can be operated with a lower voltage bias, reducing the relative contribution of readout noise.  
		
		The primary goal of this work is to reduce the wiring impedance between the LC chip and SQUID in the dfMux architecture.  This will reduce parasitics, reduce cross-talk, improve scalability, and enable operation of low-R$_{normal}$ bolometers, which will enable lower noise and optimal mapping speeds.

\section{DESIGN AND ASSEMBLY}

	In the standard dfMux cold readout system, the LC boards are attached to the back of the detector wafer at \SI{250}{mK}, while the SQUIDs are located on the \SI{4}{K} stage.  This arrangement necessitates long wires between the LC chip and the SQUID simply because they are physically far apart.  To reduce the parasitic impedance of these wires, they are made of broadside-coupled, superconducting NbTi wires encased in Kapton.  Even with these efforts to reduce the inductance, the broadside-coupled striplines have 21 $\pm$ \SI{1}{nH} of parasitic inductance\cite{avva}.  An additional 25 $\pm$ \SI{1}{nH} of parasitic inductance comes from the rest of the system, including connectors and PC board traces.  The striplines are also delicate, expensive, require labor-intensive ultrasonic soldering, and further improvements to parasitics and scalability within the existing stripline framework would be difficult.  
	
	\subsection{PCB design, connectorization, and cabling}
	
\begin{figure} [ht]
		 	\begin{center}
		 		\begin{tabular}{c} %% tabular useful for creating an array of images 
		 			\includegraphics[width=7cm]{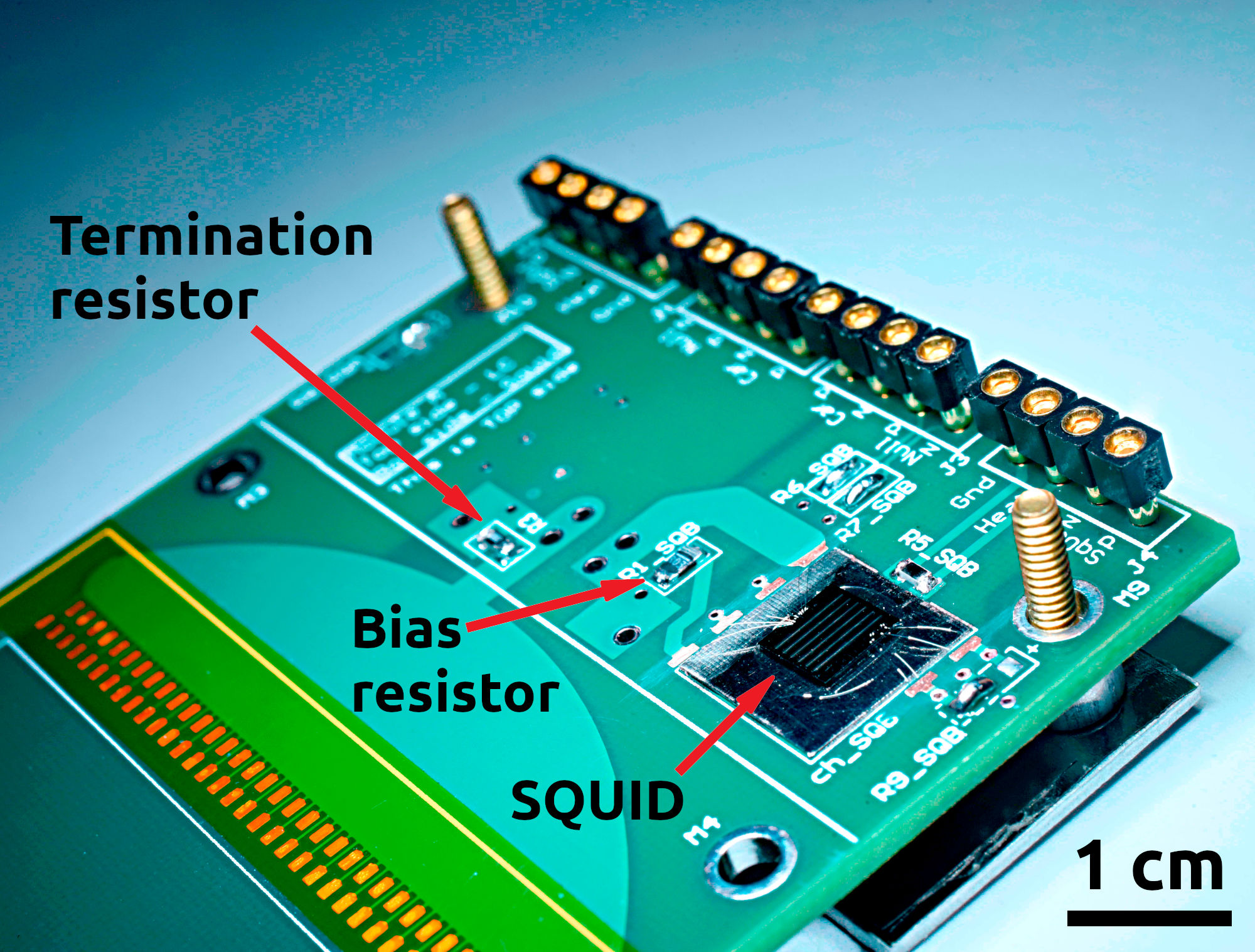}
		 			\includegraphics[width=7cm]{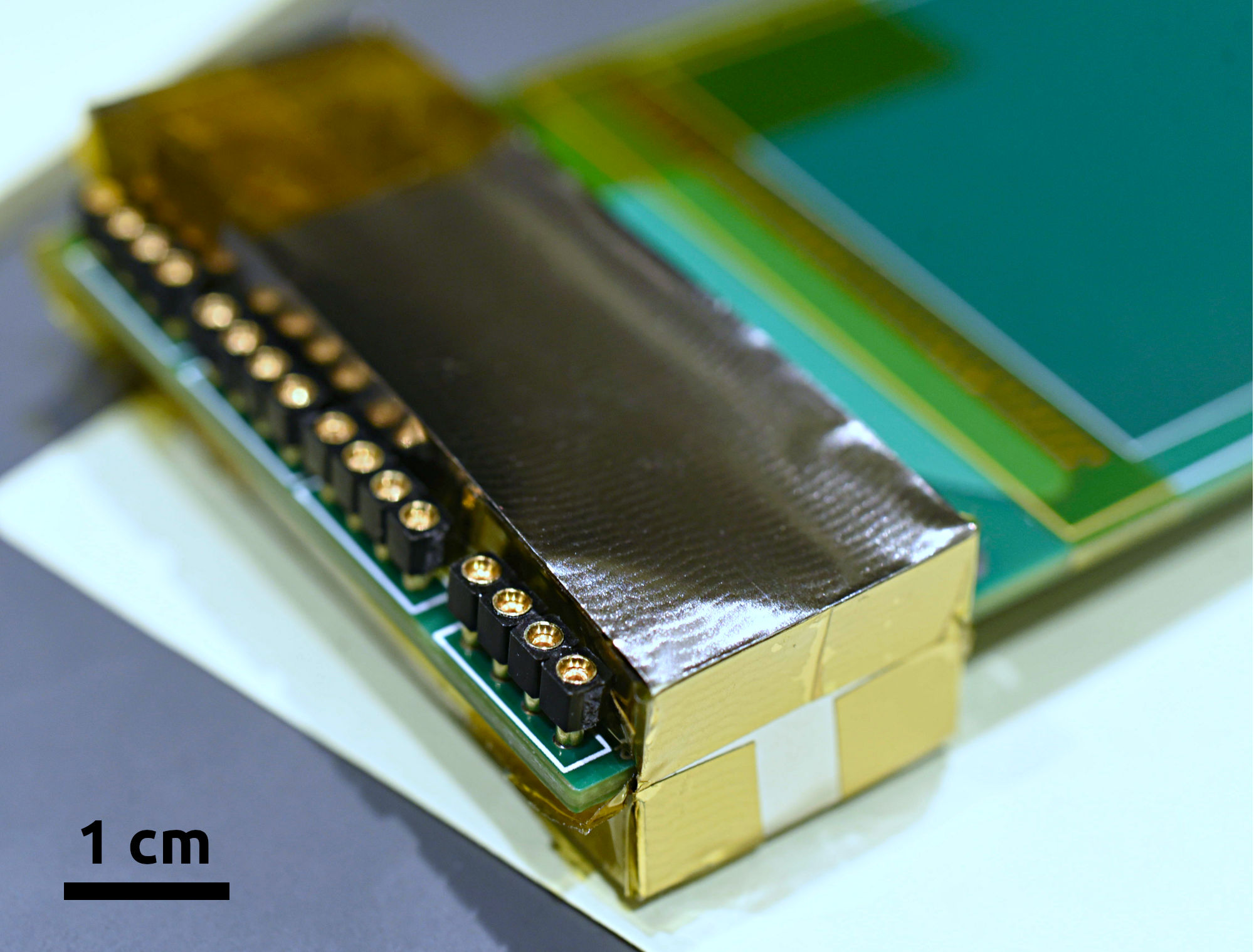}\\
                    \includegraphics[width=14cm]{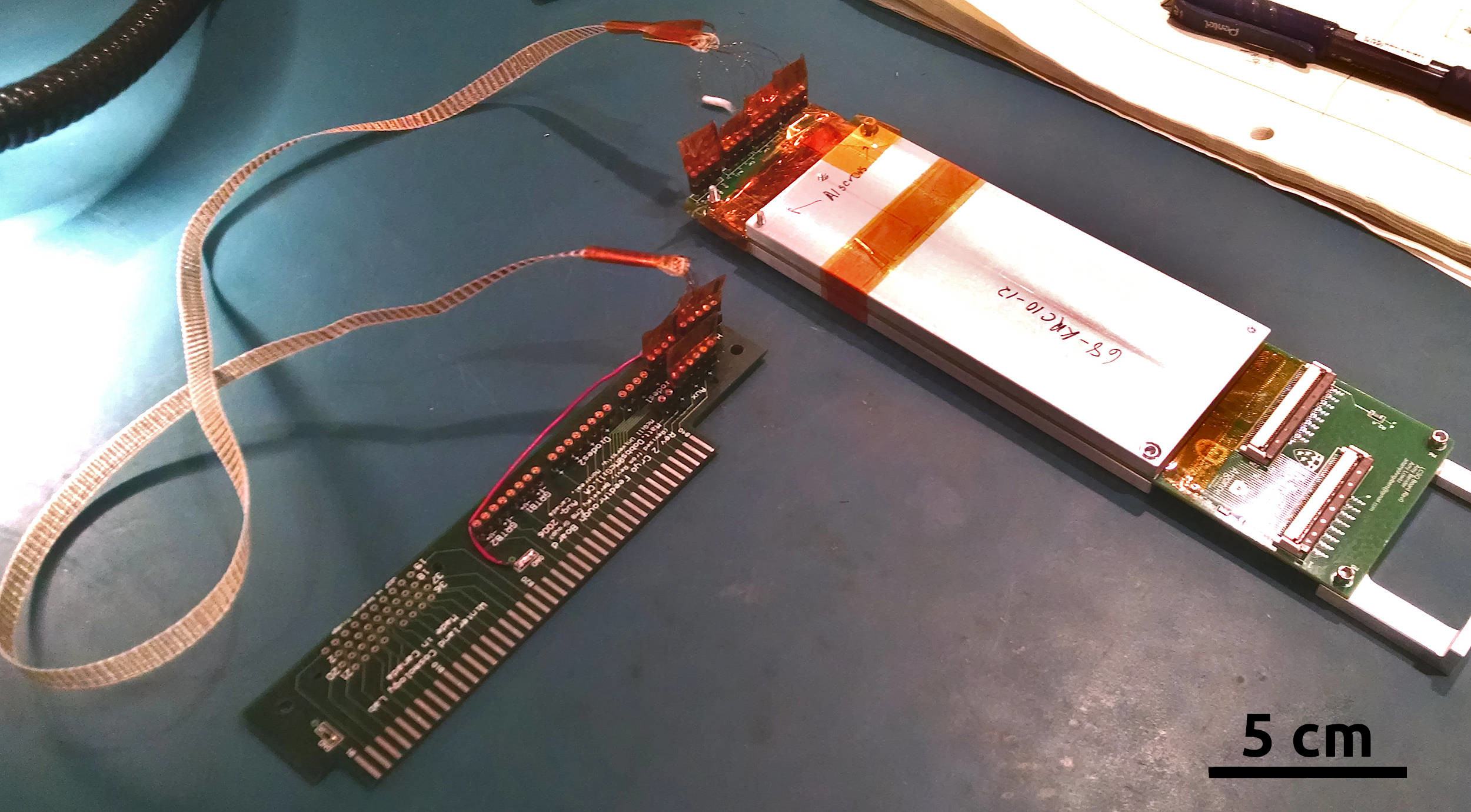}
		 		\end{tabular}
		 	\end{center}
		 	\caption[example] 
		 	{ \label{fig:pcb} 
		 		(Color online) Top left: Partially assembled LCSQ PCB without SQUID cover and without LC chip. Top right: LCSQ with protective cover and magnetic shielding installed.  Bottom: A view of the full board showing the aluminum cover for the LC chip, the NbTi ribbon cable, and the \SI{4}{K} edge connector feedthrough board which provides an interface to existing \SI{4}{K} wiring in the testbed.  
		 	}
		 \end{figure}

		In the new design, the SQUID and bias resistor are moved from the \SI{4}{K} stage to the \SI{250}{mK} stage, placed immediately adjacent to the LC chip, on the same PCB (this combined LC + SQUID PCB is referred to hereafter as an `LCSQ board').  This allows for very short PCB traces between the LC chip and the SQUID.  Where possible, the traces on the LCSQ board between the LC chip and the SQUID are wide to further reduce stray inductance.  In order to benefit as much as possible from the established dfMux cold readout architecture, most of the existing LC board design was retained.  The section of the LC board that interfaces with the detector wafer and the section that interfaces with the LC chip were unchanged.  A small extension was added to the end of the LC board, where the stripline would attach in the standard dfMux design.  On this extension are two SQUIDs (one on each side of the board, each servicing a different LC chip), wirebonding pads to allow for connections to the SQUIDs, solder pads for placement of the bias resistor, holes for connectorization, and all of the traces necessary to connect these components.  An effort has been made to keep the SQUID extension to the PCB as short as possible.  In addition to keeping on-board traces short and parasitics to a minimum, this allows the LCSQ board to fit as a drop-in replacement for standard LC boards in all of the existing SPT-3G testbeds and in the SPT-3G camera itself, which enables low-overhead laboratory testing and straightforward future on-sky demonstration.  Figure \ref{fig:pcb} shows a partially assembled LCSQ board.  
		
		To make electrical connections between the SQUID output at \SI{250}{mK} and a feedthrough board at \SI{4}{K}, 24-inch long superconducting NbTi twisted-pair cryogenic woven loom ribbon cable is used.  Because the NbTi twisted-pair cabling is after the SQUID, rather than between the LC chip and the SQUID, the relatively high stray inductance of twisted pair cabling does not contribute to the inductance budget that drives cross-talk.  The prototype LCSQ board is connectorized with single-row Samtec dip probe connectors\cite{samtec}, however future revisions will incorporate a more robust connector.

	\subsection{Magnetic shielding}
	\label{sec:mag}
	
		SQUIDs are extremely sensitive magnetometers; their performance as amplifiers can be substantially degraded if they are subject to stray ambient magnetic fields.  As a result, it is necessary to magnetically shield the SQUIDs.  In the standard dfMux architecture, two types of magnetic shielding are used.  First, the SQUIDs are housed inside an Amuneal A4K\cite{amuneal} magnetic shield.  Amuneal A4K is a high magnetic permeability material which helps to reduce the magnitude of magnetic fields near the SQUIDs.  Additionally, the SQUID is mounted on a superconducting Nb foil, which helps to `pin' any remaining field lines.  In the LCSQ design, the same Nb foil system is used to pin field lines, but instead of A4K shields, a box made from six layers of Metglas 2714A\cite{metglas} is used.  Metglas 2714A is a high magnetic permeability foil which is less costly than A4K and can be cut and folded with basic hand tools.  No significant difference in transimpedance was observed for SQUIDs shielded with Nb foil and A4K compared to SQUIDs shielded with Nb foil and the six-layer Metglas box.  Figure \ref{fig:pcb} (top right) shows the LCSQ board assembled with the Metglas box installed.

\section{MEASUREMENTS AND PERFORMANCE}
	\subsection{SQUID performance at 250 mK}
    \label{sec:SQUID}
		Eight SQUIDs have been tested with the sub-Kelvin dfMux cold readout architecture described in this work, including six NIST SA-13 SQUID arrays, one NIST SA-4 SQUID array, and one Star Cryoelectronics E2 SQUID array (referred to hereafter as SA-13s, SA-4s and E2s, respectively).\cite{sa4, starCryo}  It is important to note that none of these SQUID arrays were optimized for use below 1 K.  Nonetheless, we have seen good performance with these SQUID arrays at \SI{250}{mK}.  We plan to test SQUID arrays designed for sub-Kelvin operation with this readout architecture in the future.  
		
\begin{figure} [ht]
			\begin{center}
				\begin{tabular}{c} %% tabular useful for creating an array of images 
					\includegraphics[width=\textwidth]{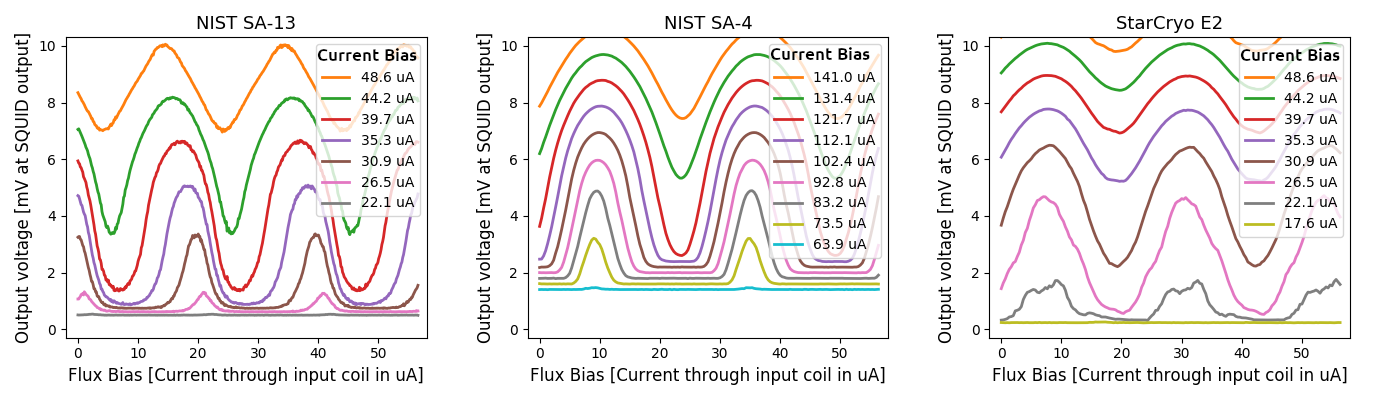}
				\end{tabular}
			\end{center}
			\caption[example] 
			{ \label{fig:vPhi} 
				Representative V$\mathrm{\phi}$ curves for NIST SA-13, NIST SA-4, and Star Cryoelectronics E2 SQUID arrays for a range of current biases and flux biases.  
                
			}
		\end{figure}

		One useful diagnostic of SQUID performance is the so-called `V$\mathrm{\phi}$' measurement: at some constant SQUID current bias, the SQUID output voltage is measured for a range of flux biases.  This measurement should yield a peak or peaks (depending on the flux bias range) with a period related to the flux quantum \cite{SQUIDtext}.  Figure \ref{fig:vPhi} shows V$\mathrm{\phi}$ curves over a range of current biases for a representative SA-13, SA-4, and E2 SQUID arrays.  All three show relatively smooth V$\mathrm{\phi}$s over a wide range of current- and flux- biases, with increased transimpedance and increased peak-to-peak variation in the SQUID output voltage at a fixed current bias (hereafter called `V$\mathrm{\phi}$ peak-to-peak') compared with their performance at \SI{4}{K}.  All of the SQUIDs tested with this architecture tuned successfully.  For comparison, a representative SA-13 tuned with a transimpedance of \SI{1138}{\Omega} and a V$\mathrm{\phi}$ peak-to-peak of \SI{5.37}{mV} at \SI{250}{mK}, and the same SA-13 tuned with a transimpedance of \SI{738}{\Omega} and a V$\mathrm{\phi}$ peak-to-peak of 4.34 mV at \SI{4}{K}.  This performance, at both \SI{4}{K} and \SI{250}{mK}, is similar to other SA-13s from the same wafer.  Based on tests of six SA-13s, there is a roughly 25\% improvement in V$\mathrm{\phi}$ peak-to-peak and a roughly 60\% improvement in transimpedance at \SI{250}{mK} compared to \SI{4}{K}.  The single E2 for which test results are currently complete shows improvement in the V$\mathrm{\phi}$ peak-to-peak and transimpedance similar to the SA-13s.  The single SA-4 for which testing has been completed shows a larger performance improvement.  At \SI{250}{mK} the SA-4 has been measured with transimpedance over \SI{1700}{\Omega} and V$\mathrm{\phi}$ peak-to-peak of \SI{5.51}{mV}.  For comparison, at \SI{4}{K} a typical SA-4 tunes with transimpedance around \SI{400}{\Omega} and V$\mathrm{\phi}$ peak-to-peak of about \SI{4.7}{mV}.

	\subsection{Bolometer operation with 250 mK SQUIDs}
	\label{sec:bolo}

\begin{figure} [ht]
		\begin{center}
			\begin{tabular}{c} %% tabular useful for creating an array of images 
				\includegraphics[width=12cm]{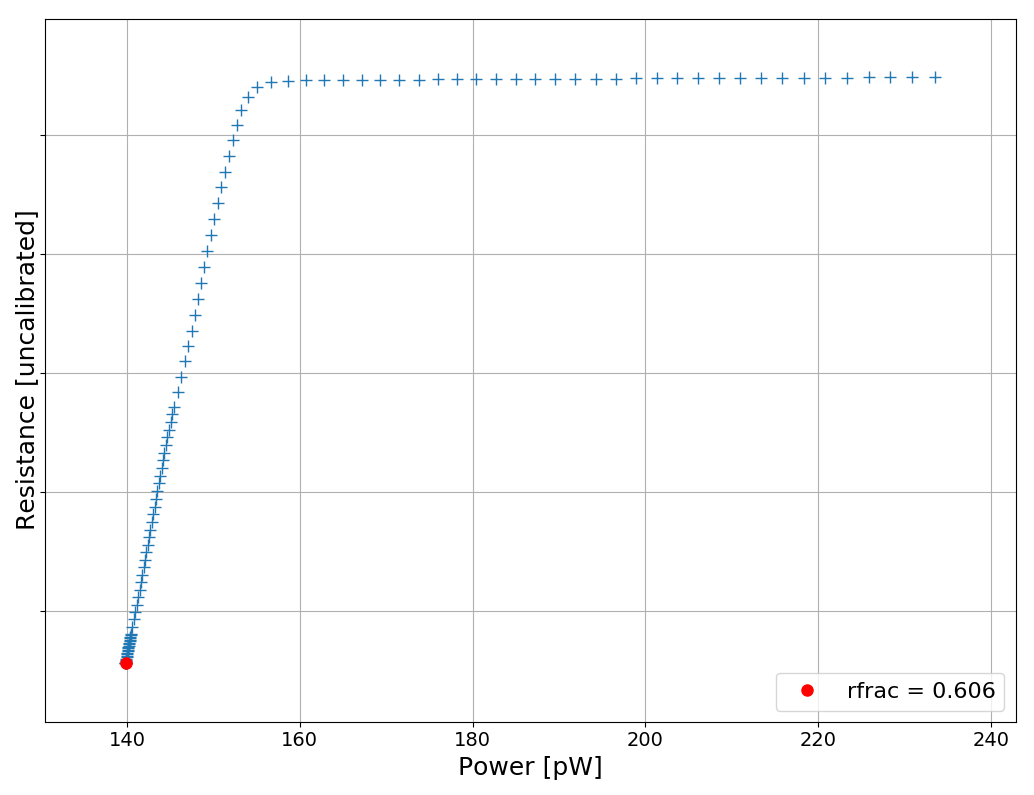}
			\end{tabular}
		\end{center}
		\caption[example] 
		{ \label{fig:bolo} 
			A representative R-P curve for a bolometer from an SPT-3G detector wafer dropping into the superconducting transition, operated with a \SI{250}{mK} NIST SA-13 SQUID at \SI{250}{mK} using the LCSQ dfMux architecture. The red dot indicates the final resistance and power of the bolometer, with a final resistance 60\% of R$_{normal}$ (rFrac=0.6). This bolometer was operated in a comb of 47 bolometers. Units on the vertical axis are uncalibrated due to the cold transfer function issue discussed in section \ref{sec:resTerm}. 
		}
	\end{figure}

	Operation of bolometers in the superconducting transition with a \SI{250}{mK} NIST SA-13 in the LCSQ architecture has been demonstrated with combs of up to 47 bolometers.  Figure \ref{fig:bolo} shows a representative resistance vs. power curve for a bolometer from an SPT-3G detector wafer being operated with the LCSQ architecture in the laboratory.  Operation of full 64-bolometer combs has not yet been demonstrated.  The current limitation to 47 bolometers comes from two sources in roughly equal parts.  Detector wafer-level flaws make a few bolometers on the comb being tested inoperable due to shorted or open TES connections.  Additionally, an in-band resonance in the LCSQ architecture is impeding LC peak-finding for a few bolometers in the comb.  However, improved performance and better damping of the in-band resonance is expected once an accurate circuit model can be developed to inform placement of damping resistors and implementation of a cold transfer function (see section \ref{sec:resTerm}).

	\subsection{Parasitic inductance}
	\label{sec:capTerm}
\begin{figure} [ht]
			\begin{center}
				\begin{tabular}{c} %% tabular useful for creating an array of images 
					\includegraphics[height=6.5cm]{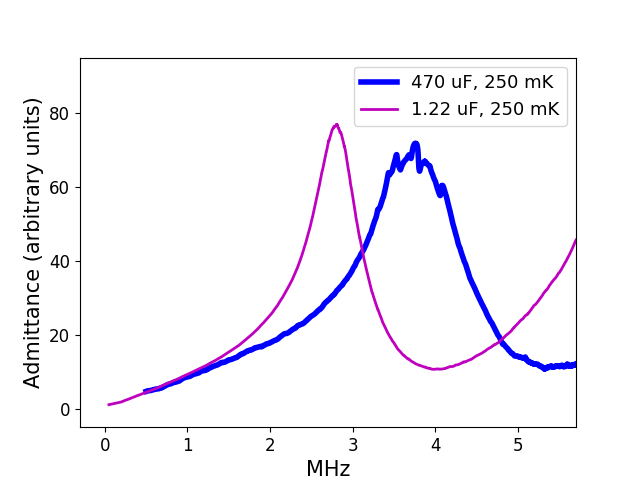}  \includegraphics[height=6.5cm]{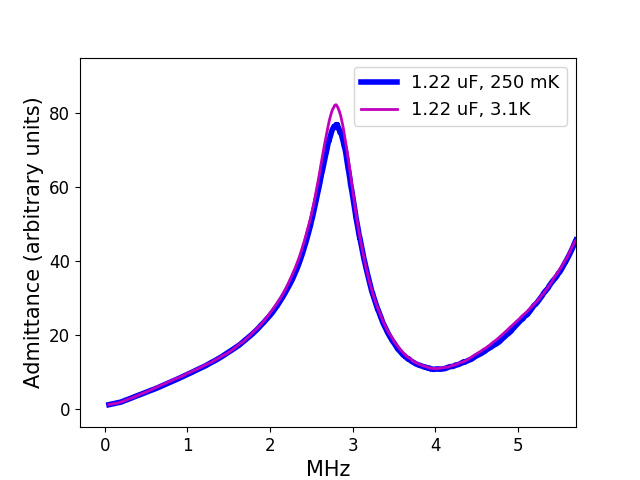}
				\end{tabular}
			\end{center}
			\caption[example] 
			{ \label{fig:capTerm}
            	Capacitor-terminated measurements of the LCSQ circuit.  Left: Measurements of the resonant peak with two different termination capacitors.  Right: Measurements taken of identical setups at two different temperatures.
			}
		\end{figure}

		Reducing the parasitic inductance of the cold readout is one of the key goals of this work, as described in section \ref{sec:motivation}.  To measure the parasitic inductance of the system, the bolometers and LC resonators are removed from the circuit and replaced with a termination capacitor between the nuller and carrier lines.  Following a similar procedure to that presented in Avva et. al. (2018)\cite{avva}, the circuit is terminated with a capacitor which is much larger than the expected stray capacitance of the system, a small AC probe tone is swept across the readout band, and the response is measured relative to the probe tone.  This procedure measures an LC resonance produced by the termination capacitor (plus any other stray capacitances) and the stray inductances.  The stray inductance is calculated from the termination capacitor value and the center frequency of the resonance according to the formula $\nu=1/2\pi \sqrt{LC}$.  With a \SI{1.22}{\mu F} termination capacitor we find a resonance centered at \SI{2.8}{MHz}, which corresponds to a parasitic inductance of \SI{2.7}{nH}.  The measurement was repeated with a \SI{470}{nF} termination capacitor and measured a resonance at 3.8 MHz, which corresponds to a parasitic inductance of \SI{3.7}{nH}.  Figure~\ref{fig:capTerm} (left) shows the results of these measurements. In the simplest circuit model, we would expect to measure the same inductance for any sufficiently large value of the termination capacitor.  The discrepancy between the \SI{1.22}{\mu F} and \SI{470}{nF} measurements suggests that the effective circuit is more complicated than anticipated.  A circuit model that captures this behavior is under active development.  For comparison, Avva et. al. (2018)\cite{avva} measured the stray inductance for the stripline cabling in the standard dfMux architecture at $\sim$\SI{20}{nH} and the stray inductance for the whole cold readout assembly of the standard dfMux architecture (including connectors, cabling, PCB traces, etc) at $\sim$\SI{45}{nH}.  Table \ref{tab:parasitics} summarizes these results.

		\begin{table}[ht]
		\caption{Parasitic inductances in the LCSQ dfMux cold readout compared to the standard dfMux readout.} 
		\label{tab:parasitics}
			\begin{center}       
				\begin{tabular}{|l|c|c|}
					\hline
					\rule[-1ex]{0pt}{3.5ex}    & cabling only & full system  \\
					\hline
					\rule[-1ex]{0pt}{3.5ex}  standard dfMux\cite{avva} & 21 $\pm$ 1 nH & 46 $\pm$ 1 nH  \\
					\hline
					\rule[-1ex]{0pt}{3.5ex}  LCSQ & - & $\leq$ 3.7 nH   \\
					\hline 
				\end{tabular}
			\end{center}
		\end{table}

		Because some of the circuit components used in the LCSQ prototype, including the SQUIDs themselves, are not optimized to work at sub-Kelvin temperatures, it was important to rule out any unanticipated low-temperature behavior of the cold circuit.  In the capacitor-terminated configuration, the response of the LCSQ circuit was measured across the readout band at \SI{3.1}{K} and at \SI{250}{mK}.  Figure \ref{fig:capTerm} (right) shows that there was no significant change is the behavior of the circuit across the band of interest between \SI{3.1}{K} and \SI{250}{mK}.

	\subsection{In-band resonance}
	\label{sec:resTerm}
	
		As discussed in section \ref{sec:bolo}, an in-band resonance is impeding peak-finding for a few bolometers in the comb.  To understand the performance of this circuit in more detail and to probe the behavior and origin of the in-band resonance, resistor terminated measurements were performed.  Similar to the capacitor-terminated measurements discussed in section \ref{sec:capTerm}, the RLC resonators are removed from the circuit and replaced with a single termination resistor.  Preliminary resistor-terminated measurements indicate that the resonance can be damped by placing appropriate resistors in certain parts of the circuit and that the NbTi twisted pair cabling plays a significant role.  Work is ongoing to develop an appropriate system of damping resistors to minimize the amplitude of the resonance and to develop circuit model that captures all of the relevant phenomenology.  In addition to informing placement of the damping resistors, an accurate circuit model will allow for calculation of a ``cold transfer function'', or mapping of ADC counts and DAC voltages to real currents in the cold circuit.  This cold transfer function can be used as a frequency-dependent calibration factor for operating SQUIDs and bolometers with appropriate DAC voltages and for accurately converting ADC counts to physical units when making measurements.  These improvements are expected to enable operation of full combs of 64 bolometers and allow for accurate measurements of noise and parasitic resistance.

\section{CONCLUSIONS}
In summary, we have demonstrated SQUID operation at \SI{250}{mK} with the sub-Kelvin SQUID architecture using NIST SA-13 SQUID arrays, a NIST SA-4 SQUID array, and a Star Cryoelectronics E2 SQUID array.  Additionally, operation of 47 TES bolometers in the superconducting transition with a \SI{250}{mK} NIST SA-13 SQUID array and been demonstrated and operation of a full comb of 64 bolometers is expected in the near future.  Significant reduction in stray inductance compared to the standard dfMux architecture has also been demonstrated.  This will enable 128x multiplexing factors by allowing for closer packing of resonators in the same readout bandwidth and will also enable operation of bolometers with lower R$_{normal}$, which require lower bias voltage and therefore have higher detector response and higher mapping speed.

Work in the immediate future will focus on damping the in-band resonance and characterizing the cold transfer function using a combination of empirical measurement and circuit modeling.  This will directly enable accurate characterization of the noise performance and parasitic resistance of this sub-Kelvin SQUID architecture.  In the longer-term, work will focus on bringing the sub-Kelvin SQUID architecture to a maturity level that will allow for deployment of a small number of LCSQ boards on-sky in a future SPT-3G observing season.

\acknowledgments % equivalent to \section*{ACKNOWLEDGMENTS}       
 
Work at the University of Chicago is supported by the National Science Foundation through grant PLR-1248097.  Work at Argonne National Lab is supported by UChicago Argonne LLC, Operator of Argonne National Laboratory (Argonne). Argonne, a U.S. Department of Energy Office of Science Laboratory, is operated under contract no. DE-AC02-06CH11357.  The McGill authors acknowledge funding from the Natural Sciences and Engineering Research Council of Canada, Canadian Institute for Advanced Research, and the Fonds de recherche du Québec Nature et technologies.

% References
\bibliography{squids} % bibliography data in report.bib
\bibliographystyle{spiebib} % makes bibtex use spiebib.bst

\end{document}